# Criteria for assessing grant applications: A systematic review

Sven E. Hug & Mirjam Aeschbach

University of Zurich

**Abstract**

Criteria are an essential component of any procedure for assessing merit. Yet, little is known about the criteria peers use to assess grant applications. In this systematic review we therefore identify and synthesize studies that examine grant peer review criteria in an empirical and inductive manner. To facilitate the synthesis, we introduce a framework that classifies what is generally referred to as 'criterion' into an *evaluated entity* (i.e. the object of evaluation) and an *evaluation criterion* (i.e. the dimension along which an entity is evaluated). In total, the synthesis includes 12 studies on grant peer review criteria. Two-thirds of these studies examine criteria in the medical and health sciences, while studies in other fields are scarce. Few studies compare criteria across different fields, and none focus on criteria for interdisciplinary research. We conducted a qualitative content analysis of the 12 studies and thereby identified 15 evaluation criteria and 30 evaluated entities as well as the relations between them. Based on a network analysis, we determined the following main relations between the identified evaluation criteria and evaluated entities. The aims and outcomes of a proposed project are assessed in terms of the evaluation criteria *originality*, *academic relevance*, and *extra-academic relevance*. The proposed research process is evaluated both on the content level (*quality*, *appropriateness*, *rigor*, *coherence/justification*) as well as on the level of description (*clarity*, *completeness*). The resources needed to implement the research process are evaluated in terms of the evaluation criterion *feasibility*. Lastly, the person and personality of the applicant are assessed from a 'psychological' (*motivation*, *traits*) and a 'sociological' (*diversity*) perspective. Furthermore, we find that some of the criteria peers use to evaluate grant applications do not conform to the fairness doctrine and the ideal of impartiality. Grant peer review could therefore be considered unfair and biased. Our findings suggest that future studies on criteria in grant peer review should focus on the applicant, include data from non-Western countries, and examine fields other than the medical and health sciences.

**Keywords**

grant peer review, grant funding, evaluation criteria, systematic review, qualitative synthesis

**Please cite this article as**

Hug, S. E., & Aeschbach, M. (2020). Criteria for assessing grant applications: A systematic review. Palgrave Communications 6(30). https://doi.org/10.1057/s41599-020-0412-9

# Introduction

Criteria are an essential component of any procedure for assessing merit (Thorngate *et al.* 2009). This is widely acknowledged in the literature on grant peer review (e.g. European Science Foundation 2011, Lamont 2009). Yet, little is known about the criteria peers use in assessing grant applications (Johnson and Hermanowicz 2017, Lamont and Guetzkow 2016, van Arensbergen and van den Besselaar 2012, van Arensbergen *et al.* 2014a). The paucity of studies on criteria of peers may be due to the following reasons. First, getting access to data of funding agencies has always been and still is difficult or even impossible (Chubin and Hackett 1990, Derrick 2018, Pier *et al.* 2018). Second, there is a multitude of criteria for assessing grant applications (Langfeldt and Scordato 2016, Moghissi 2013) and analyzing this multitude beyond simple and general criteria is challenging. Third, criteria are a blind spot in research on peer review because they are either unsayable or supposedly clear to everyone (see Tissot *et al.* 2015). By 'unsayable' we mean that criteria are tacit and therefore difficult or impossible to articulate and analyze (Gulbrandsen 2000, Ochsner *et al.* 2013). Lastly, research on peer review is mainly interested in reliability, fairness, and predictive validity (Reinhart 2012, Sabaj Meruane *et al.* 2016) and not in the criteria peers use or deem appropriate. In other words, research on peer review has ignored content validity, that is, the question of which criteria are appropriate for assessing grant applications (for a comprehensive definition of content validity, see Haynes *et al.* 1995).

However, studying and understanding the criteria that peers use to assess grant applications is crucial for a variety of reasons. For instance, according to van Arensbergen and van den Besselaar (2012), it is important to study the criteria that peers apply to improve transparency, quality, and legitimacy of grant allocation practices. Moreover, some measures suggested to improve grant peer review, both by funding agencies and in the literature, require knowledge of peers' criteria. For example, a global survey of scholars finds that 'greater training and explicit guidelines for peer reviewers are needed to ensure the quality and consistency of grant funding decisions' (Publons 2019, p. 4). Furthermore, policy makers and funding agencies recommend standardizing criteria to make grant review more efficient (NWO 2017a) and less burdensome (OECD 2018). Others suggest that criteria should be clearly defined to make grant review more reliable and less subjective (Abdoul *et al.* 2012). In addition to reasons directly related to improving grant peer review, there are also less applied reasons for studying the criteria of peers. For example, criteria can give insight into the recognition and reward system of science (Chase 1970) and the interplay between research and research policy (Langfeldt *et al.* 2019). Research on criteria can also examine presumptions such as grant peer review is 'probably anti-innovation' (Guthrie *et al.* 2018, p. 4), interdisciplinary research may be disadvantaged because 'interdisciplinary proposal reviews may have to combine multiple distinct understandings of quality' (Guthrie *et al.* 2018, p. 6), and the high degree of concentration of grant funding on certain topics and researchers may be driven by uniform assessment criteria (Aagaard 2019). Furthermore, research on peers' criteria can generate evidence on the content validity of peer review.[1] Interestingly, content validity has not been addressed in research on grant peer review although content validity is part of the paradigm that, from our point of view, implicitly underlies most of the research on peer review.[2] It seems, however, that funding agencies are becoming aware of the lack of evidence on content validity as funders from 25 countries recently concluded that 'agencies around the world use very different criteria […] in order to assess research proposals. Strikingly few of these criteria […] are evidence-based' (NWO 2017b, p. 12).

As literature reviews and compendia do not mention or only briefly discuss grant review criteria (Moghissi *et al.* 2013; for further evidence see Guthrie *et al.* 2018, Guthrie *et al.* 2019, Shepherd *et al.* 2018), we present a systematic review on this topic. In particular, we first introduce an analytical framework for structuring criteria, then identify and characterize studies on grant review criteria of peers, and eventually synthesize the criteria contained in these studies.[3] In this way, this paper contributes to advancing applied and basic research on peer review and provides a basis for empirical and theoretical research on peer review criteria, a much neglected but important topic.

Analytical framework

Peer review criteria exist in a myriad of forms and contents. This is a major challenge for analyzing criteria and, according to Langfeldt (2001, p. 822), one of the major problems for peer review researchers: 'The main characteristic of peer review – that quality criteria have no standard operationalization […] – is the main problem for students of peer review'. To address this issue, we propose a framework for structuring criteria, which is based on Scriven's *Logic of Evaluation* (1980) and Goertz' *Social Science Concepts* (2006).

According to Scriven (1980), evaluation involves the following four steps. First, the dimensions (the criteria of merit) along which the object being evaluated (the evaluand) must do well, are specified. Second, the levels of performance (the standards of merit) that indicate how well the evaluand does on a dimension are defined. Third, the performance of the evaluand is determined by comparing the evaluand to the standards of each dimension. And lastly, the results of these comparisons are synthesized into a statement of overall worth or merit. These four steps underlie all evaluation processes (Fournier 1995, Shadish 1989). We base our framework on Scriven's first three steps, as they are sufficient to analyze individual peer review criteria. From these three steps we derive the following four components of the framework. (a) The *evaluated entity* denotes the entity or object that is being evaluated. It may be the grant application as a whole or parts thereof, such as the research question, the CV of the applicant, or features of the applicant, such as her/his past performance. The evaluated entity is also called evaluand or target of the evaluation. (b) The *evaluation criteria* are the dimensions along which an entity is evaluated. Grant applications are typically assessed in terms of originality, relevance, soundness, and feasibility. Evaluation criteria are also called qualities, attributes or dimensions. According to Davidson (2005, p. 91), criteria 'distinguish a more meritorious or valuable evaluand from one that is less meritorious or valuable'. Criteria and entities can be used to generate evaluative questions such as 'Is the project (evaluated entity) innovative (evaluation criterion)?' or 'How innovative is the project? Is project X more innovative than project Y?'. It is worth noting that the term 'criterion' is inconsistently applied in the literature on peer review. It is used to denote the evaluation criterion (e.g. appropriateness), the evaluated entity (e.g. research design), or a connection of a criterion to an entity (e.g. appropriateness of the research design). In this article, we distinguish between the evaluated entity and the evaluation criterion as indicated above where analytical precision is required but otherwise we use criterion as a catch-all term. (c) The third component of the framework is the *frame of reference* of an evaluation criterion. It is a benchmark against which an entity is compared and indicates the value of an entity on a given evaluation criterion. It corresponds to Scriven's 'standards of merit' and may be metric or ordinal (including binary categories, such as 'sufficient – insufficient'). For example, higher education institutions will be graded on a 5-point scale ranging from 'unclassified' (quality that falls below the standard of nationally recognized work) to 'four star' (quality that is world-leading) in the REF 2021 (Research England *et al.* 2018, p. 101). We choose the psychological term 'frame of reference' instead of standard or benchmark to emphasize the subjective and context-dependent nature of peer review. (d) The last component of the framework is the *assigned value*. It is the value an entity achieves on a frame of reference of an evaluation criterion. For example, a journal article submitted to the REF 2021 would be assigned the value 'four star' if it were world-leading in terms of quality.

Goertz (2006, p. 6) argues that 'most important concepts [in the social sciences] are multidimensional and multilevel in nature' and, hence, can be dissected and analyzed in terms of '(1) how many levels they have, (2) how many dimensions each level has, and (3) what the substantive content of each of the dimensions at each level is.' We apply Goertz' *levels* and *dimensions* to structure the conceptual depth and breadth of the four components in the framework and *content* to denote the meaning of the components and their dimensions. In the Supplementary Materials (Part B), an example is provided of how the framework can be used analytically and it is explained how the framework can unify different structuring principles and terminologies that studies on peer review criteria have used.

Based on the framework outlined above, the research questions of this systematic review can be specified as follows: (1) What entities are being evaluated in grant peer review? (2) What criteria are used? (3) Which entities are evaluated according to which criteria?

The remainder of this article is organized into two main parts, research map and research synthesis, which according to Gough (2007, 2015) constitute a systematic review. In the first part, the research map, we delineate the inclusion criteria, the search terms, the literature search and screening, and the characteristics of the included studies. In the second part, we present the qualitative synthesis, which comprises a qualitative content analysis of the evaluation criteria and evaluated entities extracted from the included studies, a network analysis to examine the association between evaluation criteria and evaluated entities, and a similarity analysis to determine the overlap of the included studies with regard to the evaluation criteria. On the basis of our findings in both the qualitative content analysis and the network analysis, we arrive at an overall conceptualization of evaluation criteria and evaluated entities used in grant peer review. The individual steps of this systematic review were guided by the ENTREQ statement (Tong *et al.* 2012), a framework for conducting and reporting qualitative syntheses.

## Mapping the research on criteria in grant peer review

### Inclusion criteria

Studies were included if they (1) developed peer review criteria for grant proposals or established reasons used by peers for accepting or rejecting grant proposals (2) in an empirical and inductive manner, (3) reported method as well as sample (definition and size), and (4) named the examined criteria or acceptance/rejection reasons. Studies were excluded if they applied purely theoretically determined or otherwise predefined criteria (e.g. Cuca 1983, Fuller *et al.* 1991, Gregorius *et al.* 2018, Hemlin *et al.* 1995, Hume *et al.* 2015, Kaatz *et al.* 2015, Langfeldt 2001, Oortwijn *et al.* 2002, van den Besselaar *et al.* 2018, van den Broucke *et al.*

2012), if they were presumably conducted inductively, but did not specify this in their methodological approach (e.g. Agarwal *et al.* 2006, Allen 1960, Bootzin *et al.* 1992, Meierhofer 1983, Moore 1961, Porter and Rossini 1985), or if they focused on research quality in general without specifically focusing on grant peer review (e.g. Andersen 2013, Gulbrandsen 2000, Hemlin and Montgomery 1990, Hug *et al.* 2013, Mårtensson *et al.* 2016, Prpić and Šuljok 2009). Studies in which evaluation criteria were not clearly identifiable as such were also excluded (e.g. Coveney *et al.* 2017, Mow 2011).

Search terms

As a simple search with the terms 'peer review, grant, criteria' yielded less than 100 records each in Web of Science and Scopus, a more sophisticated search strategy was adopted. Particularly, search terms were established and organized in a five-step process. First, literature gathered for previous projects conducted by the research team was searched for articles that fulfilled the first inclusion criterion. Second, the references cited by these studies were screened according to the first inclusion criterion. Third, based on the studies obtained in this initial search ($n = 12$), a bibliogram (White 2005, White 2016) was prepared using VOSviewer 1.6.4 (Van Eck and Waltman 2010). In this bibliogram, words used in the title and abstracts of the identified studies were ranked by frequency. Fourth, search terms were extracted from the bibliogram and supplementary terms were identified by the research team. In this process, the following search terms were identified and organized into two categories. The terms in category A qualify the terms in category B, which indicate the subject matter relevant to this search:

- A) Assess*, evaluat*, review*, criteri*, reject*, *approve
- B) Research/grant/funding proposal*, grant applica*/allocat*/panel*/peer review, funding decision, research*/scien*/academ* funding, PROJECT FUNDING, PROJECT SELECTION, FELLOWSHIP

Lastly, search strings were created. The search terms within each category were combined with the Boolean operator 'OR' and the two categories were linked with the Boolean operator 'AND'. Capitalized terms have additionally been modified in order to restrict the outcome of the search string to the field of research funding by only searching for those terms, for example 'fellowship', in the proximity of the terms 'research*, scien*, academ*'. The proximity operator 'NEAR' was used to create search strings (e.g. fellowship NEAR academ*) in the Web of Science and the operator 'W/15' in Scopus (e.g. fellowship W/15 academ*). The two operators are equivalent. The full search strings are listed in the Supplementary Materials (Part C).

## Searching and screening

The search strategy was English-language based, but publications written in further languages understood by the research team (French, German, Italian, and Spanish) were screened as well. Grey literature has not been searched systematically, however, grey literature identified in the searches was included in this article. The search and screening process was carried out between August and October 2018.

Using the search strings indicated above, the Web of Science and Scopus were searched and records screened in several rounds (Figure 1). In case of uncertainty, publications were included in the next screening round. In the first round, all titles were screened to determine publications relevant to criteria for grant peer review (i.e. inclusion criterion 1). As the search strategy favours inclusion over specificity, it yielded 12,628 publications and it was possible to exclude as many as 11,723 publications on the basis of the title. In the second round, the abstracts of 905 publications were screened according to inclusion criteria 1 and 2 and 800 publications that did not meet these two criteria were excluded. In the third round, the full texts of 105 publications were examined and those that clearly did not meet criteria 1 to 4 were excluded. After this first screening of full texts, 43 publications were identified as highly likely to meet all four inclusion criteria. At this point, a citation-based search exploiting direct citation relationships (Belter 2016) was conducted in order to complement the search term based queries described above. References and citing publications of the 43 publications served as input for this search. While references were directly extracted from the publications, citing publications were searched in the Web of Science, Scopus, and Google Scholar. A total of 3,558 records were identified in this way, whereby the research team found high redundancy across databases and a large overlap with the results of the prior searches in the Web of Science and Scopus. Therefore, the titles of these records could be screened quickly and 3,541 publications were discarded based on inclusion criteria 1 and 2. The full texts of the remaining 47 publications were screened and those clearly not meeting criteria 1 to 4 were excluded ($n = 10$). An additional 37 potentially relevant studies were identified in this process. In the sixth and final screening round, the full texts of the 43 studies included after the third screening round plus the 37 studies resulting from the citation-based search were closely examined and final inclusion decisions were made on the basis of all four inclusion criteria. In total, 12 studies were included in the qualitative synthesis (i.e. Abdoul et al. 2012, Guetzkow *et al.* 2004, Hartmann and Neidhardt 1990, Lahtinen *et al.* 2005, Lamont 2009, Pier *et al.* 2018, Pollitt *et al.* 1996, Reinhart 2010, Schmitt *et al.* 2015, Thomas and Lawrence 1991, van Arensbergen *et al.* 2014b, Whaley *et al.*

2006). Hartmann's and Neidhardt's (1990) joint article was included, but not their books (Hartmann 1990, Neidhardt 1988), as they contain the same criteria as the joint article. Although the study of Guetzkow et al. (2004) is part of Lamont's book (2009) and both studies are based on the same data, the two studies were both included as they applied different structuring principles to the material. However, the criteria from Guetzkow et al. (2004) were only included once in the qualitative synthesis to avoid duplication of criteria.

**Figure 1**. Flow diagram of searching, screening, and inclusion of publications. Black indicates search processes and grey indicates screening processes.

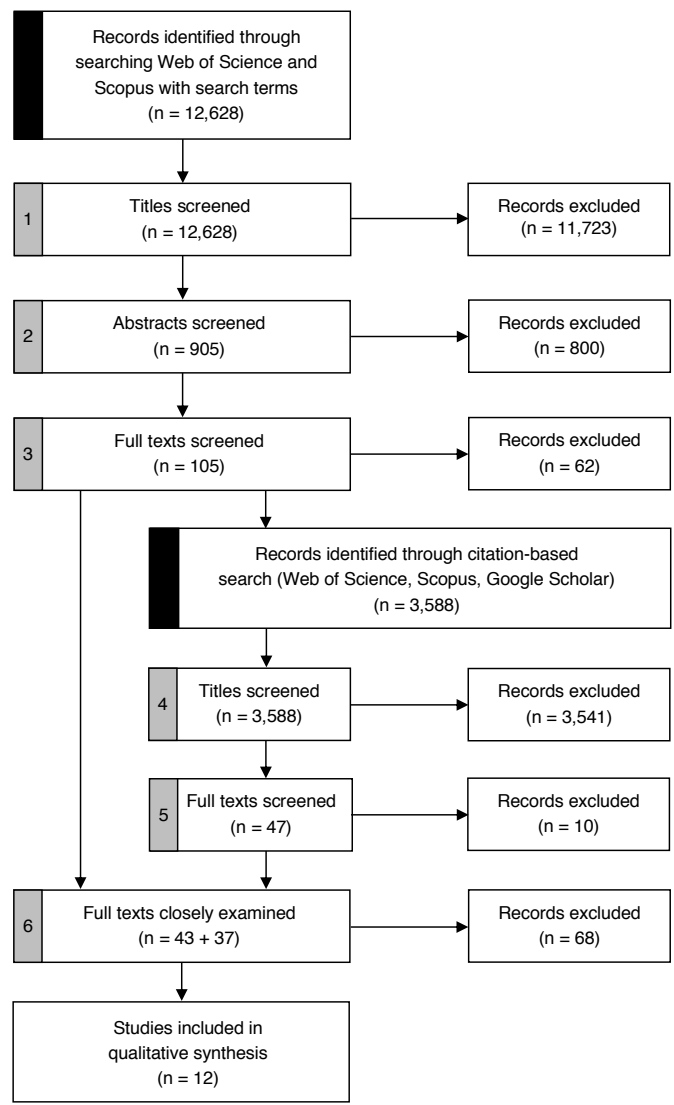

Study characteristics

The included studies were coded for the following 14 characteristics: publication year, publication language, document type of publication, field of research studied, region funding agency belongs to, type of funding agency, objective of examined funding program, purpose of study, study conducted by insider or outsider, method of data collection, stage of review process criteria refer to, sample size, criteria levels per study, and criteria dimensions per study. The *fields of research* examined in the studies were categorized according to the six broad fields of the FORD classification (OECD 2015). Based on the abstract, introduction and discussion, the *purpose of a study* was either coded as improving or understanding grant peer review. Studies with an improvement focus address deficits of peer review (e.g. low reliability), set up review criteria for a funding program, or educate potential applicants about common deficiencies of grant applications. In contrast, studies with a focus on understanding grant peer review are not interested in applied research questions and examine, for example, the criteria actually used by peers, the meaning of criteria in different disciplines, or epistemic and social aspects of the review process. To determine whether a *study was conducted by insiders* (i.e. researchers examined a grant peer review process in their own field) or outsiders, the authors' affiliation reported in the study were compared with the fields examined in the study. The *methods of data collection* were categorized according to whether data was collected from actual peer review processes (i.e. written reviews, oral comments) or whether information was elicited from scholars (i.e. through interviews, surveys, or the Delphi method). The *objectives of the examined funding programs* were coded as either funding projects (e.g. research or collaboration projects) or funding scholars (e.g. career development, scholarships). *Levels* and *dimensions* from the analytical framework were used to quantify how each study structured its criteria. While all dimensions on each level were counted, only the dimensions on the lowest level will be reported since this is the best indicator of the conceptual breadth of a study.

**Table 1**. Characteristics of studies that inductively examined grant review criteria. Data given as number and percentage of total studies included (N = 12).

| Characteristic | Summary data |
| --- | --- |
| Publication year | |
|     First study | 1990 |
|     Latest study | 2018 |
|     Mean | 2006 |
|     Median | 2007.5 |
| Fields in which criteria were studied[1] | |
|     Natural sciences | 2 (17%) |
|     Engineering and technology | 2 (17%) |
|     Medical and health sciences | 8 (67%) |
|     Agricultural sciences | 0 (0%) |
|     Social sciences | 4 (33%) |
|     Humanities | 3 (25%) |
| Region in which data was collected | |
|     USA | 6 (50%) |
|     Europe (CH, FRA, FIN, GER, NLD) | 6 (50%) |
| Purpose of study | |
|     Improving grant peer review | 7 (58%) |
|     Understanding grant peer review | 5 (42%) |
| Study conducted by | |
|     Insiders | 6 (50%) |
|     Outsiders | 6 (50%) |
| Method of data collection | |
|     Interview, survey, Delphi method | 7 (58%) |
|     Actual reviews and comments | 5 (42%) |
| Stage of review process criteria refer to | |
|     Individual review | 7 (58%) |
|     Panel review | 3 (25%) |
|     Individual and panel review | 1 (8%) |
|     Not reported | 1 (8%) |
| Objective of examined funding program | |
|     Funding projects (knowledge creation) | 6 (50%) |
|     Funding scholars (talent development) | 3 (25%) |
|     Not reported | 3 (25%) |
| Criteria levels per study ('depth') | |
|     Minimum | 1 |
|     Maximum | 4 |
|     Mean | 2.25 |
|     Mode[2] | 2 |
| Criteria dimensions per study ('breadth') | |
|     Minimum | 7 |
|     Maximum | 66 |
|     Mean | 26 |
|     Median | 21.5 |

*Note.* 1: Some studies included more than one field in their analysis. 2: As the data distribution does not allow to indicate a meaningful median, the mode is reported.

The main characteristics of the 12 included studies are summarized in Table 1. The first study on grant review criteria, which is clearly identifiable as inductive and empirical, was conducted by Hartmann and Neidhardt in 1990. The studies were published in English (n = 11) or German (n = 1) and as journal article (n = 11) or book (n =1). Two-thirds of the studies examine criteria in the medical and health sciences, while studies in other fields are scarce. Few studies compare criteria across different fields (n = 3) and none focuses on criteria for interdisciplinary research. In general, studies on criteria in the medical and health sciences were done by insiders, involve just one field or discipline, and focus on improving grant peer review. In contrast, studies on other fields were conducted by outsiders, involve two or more fields, and focus on understanding grant peer review. The included studies cover funding agencies from six countries (Finland, France, Germany, Netherlands, Switzerland, USA), which can be grouped in two regions (USA and Continental Europe). Although the studies typically focus on large government agencies (e.g. NIH, German Research Foundation), there are also three studies that analyze university-based funding schemes and small, non-governmental agencies. The objectives of the examined grant schemes were either to fund research projects (n = 6) or scholars (n = 3). Three studies did not report any objectives. Grant schemes with other funding objectives, such as collaboration, infrastructure, or knowledge transfer are not among the included studies. Five studies collected data from actual peer review processes (written reviews, oral comments) while seven elicited information from scholars. In the latter studies, scholars were either interviewed about the criteria they have used in panels and written reviews or they were involved in multi-stage designs (e.g. Delphi survey) to set up review criteria for funding programs. The sample size of the studies that used actual review data is on average larger (unit: documents, mean = 308, median = 212, minimum = 51, maximum = 639) than the sample size of the studies that elicited data from scholars (unit: persons, mean = 48, median = 48.5, minimum = 12, maximum = 81). Most studies analyzed criteria of individual reviews (n = 7), while only few studies focused on criteria used in panels (n = 3). One study based its analysis on data from both individual reviews and panels and another does not report the stage of the review process the criteria refer to. The structure of the criteria can generally be characterized as flat and broad. Nine studies organize their criteria on one or two hierarchical levels and three studies on three or four levels. The number of dimensions (i.e. the breadth of the reported criteria) ranges widely from seven to 66 and averages 26 (median = 21.5).

# Qualitative synthesis

## Methods

*Quality and relevance appraisal.* Systematic reviews usually involve an appraisal of the quality and relevance of the included studies 'to judge the usefulness of the results [of the included studies] for answering the review question' (Gough 2007, p. 219) and 'to determine how much "Weight of Evidence" should be given to the findings of a research study' (Gough 2007, p. 214). In the present study, we did not conduct such an appraisal for two reasons. First, we have applied narrow inclusion criteria and we therefore consider all included studies as useful. Second, giving some studies more weight than others would not be meaningful with regard to the research questions.

*Data extraction.* All included studies provided short descriptions of the criteria either in a table or as a list in the text. These descriptions, along with the criteria names, were entered verbatim into Microsoft Excel and the original structure of the criteria was preserved. Some studies also provided quotes or discussed criteria in great detail. Such text parts were not included as it was extremely difficult or impossible to distinguish the criterion under discussion from other criteria also contained in these text strings. A total of 312 criteria were extracted.

*Qualitative content analysis.* A qualitative content analysis (Mayring 2014, Mayring 2015) was conducted to summarize the content of the extracted data. The analytical framework outlined in the *Introduction* served as background knowledge in the content analysis. First, the extracted data was split into entity and criteria segments to disentangle complex configurations consisting of multiple entities and criteria, and to relate criteria to entities, which was necessary to answer the third research question. In particular, a new row was used for each entity in Excel and the corresponding criteria were written in separate columns in the same row. For example, 'budget and equipment are described and appropriate' was split into 'budget – described, appropriate' and 'equipment – described, appropriate'. This segmentation expanded the extracted data from 312 to 373 rows. In a second step, two coders (MA, SEH) independently developed codes for evaluated entities and evaluation criteria based on the first half of the data in Atlas.ti 8. In particular, coders went through the data line by line and generated a new code each time data could not be subsumed under existing codes. The coders then compared and discussed their codes and agreed on common codes. They worked through the first part of the data again and then jointly revised and finalized the codes. Residual codes (e.g. 'other entity') were created so that that all segments could be coded. This resulted in 30 entity codes and 15 criteria codes. In a third step, each coder independently coded the whole data in Excel, assigned one code to each entity (one variable with 30 codes), and decided which of the criteria were reported for an entity (15 variables with the two codes 'present' and 'absent'). For example, the string 'budget is described and appropriate' was assigned to 'budget' (entity code) and 'completeness – present 'as well as 'appropriateness – present' (criteria codes). After coding was completed, Krippendorff's alpha (Krippendorff 2004, Krippendorff 2011) was computed in R 3.5.2 (R Core Team 2018) using the *icr* package (Staudt 2019) to assess the inter-coder reliability of the two coders with regard to the evaluation criteria (15 variables) and the evaluated entities (one variable). Alpha was calculated for each of these variables and the coefficients of the 15 criteria variables were then averaged to provide a single reliability index for the coded criteria. Krippendorff's alpha was 0.78 for the entity variable and 0.69 for the averaged criteria variables, indicating substantial agreement according to Landis and Koch (1977). According to Krippendorff's (2004) more conservative benchmarks, these alpha values allow drawing tentative conclusions. Based on the reliability analysis, the coders discussed coding disagreements until consensus was reached.

*Conceptual counting*. The frequencies of the coded entity and criteria variables were computed and transformed from 'full counting' to 'conceptual counting' to balance quantitative peculiarities of individual studies. While full counting takes into account every occurrence of a code, conceptual counting considers an occurrence of an unconnected code (i.e. an entity without a related criterion, a criterion without a related entity) or a combination of codes (i.e. a particular connection of a criterion to an entity) only once per study. For example, the connection of 'extra-academic relevance' to 'project in general' occurred 19 times in Schmitt et al. (2015) but was considered only once in conceptual counting.

*Jaccard Index*. Following (Fried 2017), the Jaccard Index, a similarity measure for binary data, was calculated to determine the overlap of the included studies with regard to the coded evaluation criteria. The Jaccard Index was defined as the number of criteria two studies share divided by the sum of shared criteria and the criteria unique to each of the two studies. The publications of Guetzkow et al. (2004) and Lamont (2009) were processed as one study because they are complementary in terms of criteria. The Jaccard Index was calculated for each study pair and the indices were then averaged to provide a single overlap measure. A Jaccard Index of 0 indicates no overlap and 1 indicates total overlap.

*Network analysis*. To examine the association between evaluation criteria and evaluated entities in terms of content, a bipartite network was created in VOSviewer 1.6.4 (Van Eck and Waltman 2010) with two types of nodes (evaluated entities, evaluation criteria) and connections between entities and criteria as edges. Node sizes and edge weights were taken from the conceptual counting data. The residual categories 'no entity reported' and 'other entity' (see Table 2) were not included in the network as they have no relevant semantic meaning. The entity 'social skills' was also not included as it had no connection to a criterion. To detect clusters of criteria and entities in the network, the DIRTLPAwb+ algorithm (Beckett 2016) implemented in version 2.11 of the R package *bipartite* (Dormann *et al.* 2008) was used. The analysis was run with the function *metaComputeModules* using the default parameters. In network science, clusters are called communities. Hence, the detected clusters are indicated as communities in this article. Lastly, the following descriptive statistics were calculated to characterize the association between criteria and entities quantitatively: number of criteria which were reported without a related entity, number of entities which were reported without a related criterion, average number of reported criteria per reported entity, number of different entities to which a criterion was connected.

Results

*Evaluated entities*. The entities were organized on two levels and specified by four dimensions on the basic level and by 30 dimensions on the second level in the qualitative content analysis (Table 2). The basic level comprises the four entities applicant, project, environment, and other. The residual category 'other entity' consists of items that could be of interest when studying grant peer review processes holistically (e.g. statements about the reviewer herself/himself), but it does not contain any entity referring to grant applications. Hence, at the basic level, applicant, project, and environment can exhaustively describe the evaluated entities of grant applications. Quantitatively, the project is the most important entity by far (72% of the assigned codes; full counting, N = 373) while the applicant plays a minor role (21%) and the environment is marginal altogether (3%). Correspondingly, the project is the most detailed entity and comprises 19 sub-entities, while the applicant consists of eight sub-entities, and the environment includes just one sub-entity. Moreover, both applicant and project feature a residual category ('other'), indicating that there are additional sub-entities relevant to the evaluation of grant applications, which are not listed in Table 2. To reflect that the included studies reported evaluated entities at different levels of abstraction, four sub-entities were defined in more general terms in the qualitative content analysis (i.e. applicant in general, generic qualifications, project in general, method in general). They account for 26% of the totally assigned entity codes (full counting, N = 373). Since the content of the evaluated entities will be jointly interpreted with the evaluation criteria in the section *Association between criteria and entities*, the content of the entities is not discussed here.

**Table 2**. Evaluated entities resulting from the qualitative content analysis.

| Entity | Description | Full counting (N = 373) | Conceptual counting (N = 256) |
|---|---|---|---|
| **Applicant** | | 80 (21%) | 47 (18%) |
| Applicant in general | Entity is concerned with the investigator(s), applicant(s), co-applicant(s), collaborator(s), and the research team in general without providing further details.<br>*Example*: 'evaluation of the team (collaborators, consultants)' | 26 (7%) | 7 (3%) |
| Generic qualifications | Entity is concerned with professional or expert qualifications of the applicant(s) in general without providing further details.<br>*Example*: 'The qualifications of the staff/personnel are adequate to meet project's goals.' | 6 (2%) | 5 (2%) |
| Research skills | Entity is concerned with the knowledge, expertise, and research skills of the applicant.<br>*Examples*: 'broad expertise', 'writing skill', 'investigator not familiar with particular data base or technique' | 11 (3%) | 9 (4%) |
| Social skills | Entity is concerned with social skills of the applicant.<br>*Examples*: 'ability to motivate others', 'fit in a group', 'leadership skills' | 7 (2%) | 1 (1%) |
| Academic background | Entity is concerned with the training, education and professional experience the applicant has made, with previous/current employers, and with former/current positions.<br>*Examples*: 'past experience of the applicant', 'previous employers/institute' | 15 (4%) | 11 (4%) |
| Past performance | Entity is concerned with the applicant's past research performance, research accomplishments, previous publications, and grants.<br>*Examples*: 'track record of the applicant', 'insufficient professional publications' | 7 (2%) | 6 (2%) |
| Reputation | Entity is concerned with the reputation, esteem, and prestige of the applicant.<br>*Examples*: 'applicant's esteem within the scientific community', 'awards' | 4 (1%) | 4 (2%) |
| Other | Entity is concerned with aspects related to the applicant which could not be assigned to any of the other applicant entities.<br>*Example*: 'researcher time fully scheduled' | 4 (1%) | 4 (2%) |
| **Project** | | 269 (72%) | 191 (75%) |
| Project in general | Entity is concerned with the proposed research project in general and is referred to as (proposed) study, project, research, or as application, proposal.<br>*Example*: 'originality of the study' | 50 (13%) | 29 (11%) |
| Current state | Entity is concerned with the current state of research and the literature review.<br>*Example*: 'Literature review inadequate or inappropriate.' | 3 (1%) | 3 (1%) |

| Entity | Description | | |
|---|---|---|---|
| Topic | Entity is concerned with the research topic and the content of the proposed study.<br>*Examples*: 'new topic', 'The health care concept to be examined is tailored to patient needs and equal opportunities' | 15 (4%) | 6 (2%) |
| Research question | Entity is concerned with the research question or problem, the hypotheses, the scope or focus of the study, and the research goals or aims.<br>*Example*: 'The research question is relevant for patients.' | 19 (5%) | 17 (7%) |
| Theory | Entity is concerned with theoretical and conceptual aspects of the proposed research.<br>*Example*: 'The project is guided by a clear theoretical framework, model, or philosophy of mental health.' | 15 (4%) | 9 (4%) |
| Approach | Entity is concerned with the approach of the proposed research.<br>*Example*: 'original approach' | 10 (3%) | 4 (2%) |
| Preparatory work | Entity is concerned with preparatory work that is directly related to the proposed project.<br>*Example*: 'Pilot work not done, or pilot results not adequately discussed or conflict with proposal.' | 3 (1%) | 3 (1%) |
| Data | Entity is concerned with the data or sample, its properties, and data collection and handling.<br>*Example*: 'data collection and/or data management procedures unclear, inappropriate, or unreliable' | 25 (7%) | 22 (9%) |
| Ethics | Entity is concerned with ethical aspects of the proposed research such as implications for participants or the independence of applicants from sponsors.<br>*Examples*: 'lack of medical supervision', 'There is no promotion of industrial interests.' | 11 (3%) | 7 (3%) |
| Method in general | Entity is concerned with method(s) and methodology in general without providing further details.<br>*Examples*: 'synthesis of methods', 'deficiency in methodology' | 16 (4%) | 11 (4%) |
| Methodological details | Entity comprises a wide variety of methodological details.<br>*Examples*: 'interviewer standardization (training) not described', 'instrument psychometric properties not established' | 18 (5%) | 13 (5%) |
| Research design | Entity is concerned with the research design.<br>*Example*: 'research design problems' | 9 (2%) | 9 (4%) |
| Evaluation | Entity is concerned with the quality assurance, monitoring and evaluation of the proposed research processes and the evaluation of the outcomes of the project.<br>*Examples*: 'The concept for quality assurance and quality management of the project is described in the application.', 'The evaluation includes patient-relevant endpoints wherever possible.' | 13 (3%) | 7 (3%) |
| Analysis | Entity is concerned with the analysis and the analytical plan of the proposed research.<br>*Examples*: 'inappropriate statistical analysis', 'analytic plan lacks detail or justification' | 8 (2%) | 7 (3%) |
| Results | Entity is concerned with the anticipated results of the project and is indicated by nouns such as outcome, impact, consequences, discoveries, contributions, insights, findings, improvements, understanding, and knowledge. | 22 (6%) | 16 (6%) |
| | *Example*: 'The project makes an innovative contribution to the field of mental health.' | | |
| Budget | Entity is concerned with the requested financial resources, costs, budget and budget plan.<br>*Example*: 'Comment on the application's budget' | 11 (3%) | 11 (4%) |
| Resources | Entity is concerned with required or available resources and includes references to resources in general, equipment resources, and personnel resources. References to resources explicitly linked to the research environment and financial resources are not included in this entity.<br>*Examples*: 'Resources not well described', 'The resources described are adequate to carry out the project.' | 8 (2%) | 6 (2%) |
| Project plan | Entity is concerned with the schedule and timeline as well as the general course of action of the project (e.g. the working or research plan).<br>*Example*: 'Comments on the presented working plan, the appropriateness of the timing or coordination between different research units.' | 9 (2%) | 7 (3%) |
| Other | Entity is concerned with aspects related to the project which could not be assigned to any of the other project entities.<br>*Example*: 'Weak dissemination plan' | 4 (1%) | 4 (2%) |
| **Environment** | | 11 (3%) | 9 (4%) |
| Research environment | Entity is concerned with the environment, in which the proposed research will be conducted. It refers to the institutions, in which the project will be executed, and institutional resources, such as facilities, equipment or staff, that are provided for the project.<br>*Example*: 'statements about groups, laboratories, institutes, departments, or universities where the project will be performed' | 11 (3%) | 9 (4%) |
| **Other** | | 13 (3%) | 9 (4%) |
| No entity reported | Code was applied to statements where an evaluation criterion was reported without a corresponding entity.<br>*Example*: 'originality' | 7 (2%) | 7 (3%) |
| Other entity | Code was applied to statements concerned with entities other than the applicant, the project, or the environment.<br>*Example*: 'statements about the reviewer himself or herself.' | 6 (2%) | 2 (1%) |

*Note.* Percentages do not add up to 100% due to rounding.

*Evaluation criteria*. Fifteen evaluation criteria were identified in the qualitative content analysis (Table 3). They were organized on one level and comprise a residual category ('other'), indicating that there are additional criteria relevant to the evaluation of grant applications, which are not listed in Table 3. The six most frequent criteria account for more than two-thirds of the totally assigned criteria codes (extra-academic relevance, 14%; completeness, 13%; appropriateness, 11%; originality, 10%; clarity, 10%; feasibility, 10%; full counting, N = 387). As the included studies reported evaluation criteria at different levels of abstraction, two were defined in more general terms (i.e. quality, general relevance). However, these two evaluation criteria occurred rarely (4% of all assigned criteria codes; full counting, N = 387). Since the content of the evaluation criteria will be interpreted together with the evaluated entities in the section *Association between criteria and entities*, the content of the criteria is not further addressed here.

**Table 3**. Evaluation criteria resulting from the qualitative content analysis.

| Criterion | Description | Full counting (N = 387) | Conceptual counting (N = 279) |
| --- | --- | --- | --- |
| Quality | Criterion evaluates an entity in terms of general quality (incl. quality, poor – good, weak – strong). *Examples*: 'methodological quality', 'weak dissemination plan' | 10 (3%) | 10 (4%) |
| Originality | Criterion evaluates the originality of an entity. Evaluations of originality are indicated by adjectives such as new, novel, original, innovative, unusual, unconventional, or nouns derived from these adjectives. *Examples*: 'originality of the study', 'new theory' | 40 (10%) | 20 (7%) |
| General relevance | Criterion evaluates the relevance of an entity without specifying for whom or what the entity is of value. Evaluations of relevance are indicated by nouns such as significance, relevance, importance, usefulness, timeliness, topicality, or adjectives derived from these nouns. *Example*: 'significance of the proposal's focus' | 4 (1%) | 4 (1%) |
| Academic relevance | Criterion evaluates the relevance of an entity for academia (e.g. significance for the scientific community, a research field, for scientific/theoretical advances). *Examples*: 'significance of the scientific investigation within its own field', 'significance of impact on academia' | 18 (5%) | 16 (6%) |
| Extra-academic relevance | Criterion evaluates the relevance of an entity for the non-academic sphere (e.g. for society, policy, economy, technology, education, health care). *Example*: 'the research has practical relevance for health promotion activities', 'relevance of the results for solving societal, economic, technical or psychic problem' | 54 (14%) | 22 (8%) |
| Appropriateness | Criterion evaluates the appropriateness of an entity. Evaluations of appropriateness are indicated by adjectives such as appropriate, adequate, sufficient, suitable, or nouns derived from these adjectives. *Examples*: 'appropriateness of the funds requested', 'insufficient professional publications', 'inappropriate protection of human subjects' | 44 (11%) | 38 (14%) |
| Rigor | Criterion evaluates whether/how an entity has been, is or will be done according to scholarly standards for conducting research. Evaluations of rigor are indicated by verbs (e.g. done, established, measured, estimated, studied, considered, planned, operationalized, pre-registered), adjectives (e.g. sound, rigorous, solid, unreliable, problematic), or nouns derived from these adjectives. | 26 (7%) | 20 (7%) |

| Criterion | Description | | |
|---|---|---|---|
| Coherence/ justification | *Examples*: 'important variables not measured or studied', 'pilot work not done', 'the approach is sound', 'the evaluation plan is rigorous.' Criterion evaluates the coherence of one or several entities or whether an entity is justified. Evaluations are indicated by adjectives such as aligned, coherent, compatible, connected, consistent, justified, or nouns derived from these adjectives. *Examples*: 'there is coherence between the research problem(s), research question(s) and research methodology', 'analytic plan lacks justification' | 27 (7%) | 25 (9%) |
| Completeness | Criterion evaluates whether an entity is (completely) described or reported. Evaluations are indicated by verbs such as addressed, articulated, described, defined, delineated, discussed, detailed, specified, stated, reported. *Examples*: 'All key elements of the research are defined.', 'The evaluation plan is described in the proposal.' | 51 (13%) | 41 (15%) |
| Clarity | Criterion evaluates an entity with regard to its comprehensibility and clarity. Evaluations are indicated by adjectives such as clear, comprehensible, explicit, organized, well written/articulated, or nouns and adverbs derived from these adjectives. *Examples*: 'clear presentation (interview)', 'application poorly written and/or disorganized' | 37 (10%) | 33 (12%) |
| Feasibility | Criterion evaluates the feasibility of an entity. Evaluations of feasibility are indicated by adjectives such as capable, feasible, practical, realistic, viable, or nouns derived from these adjectives. *Example*: 'Verifiable qualifications and evidence of data access demonstrate that the applicants are capable of carrying out the project.' | 37 (10%) | 31 (11%) |
| Diversity | Criterion evaluates an entity in terms of diversity and heterogeneity. *Examples*: 'institutional diversity', 'disciplinary diversity', 'cultural diversity among the organization's staff and board' | 11 (3%) | 8 (3%) |
| Motivation | Criterion evaluates the motivation of an applicant. Evaluations of motivation are indicated by nouns such as ambition, determination, perseverance, or willingness. *Example*: 'enthusiasm of the applicant' | 7 (2%) | 2 (1%) |
| Traits | Criterion evaluates an applicant along a variety of personality traits. Such evaluations are indicated by nouns such as authenticity, humility, self-consciousness, adjectives such as intelligent, independent, talented, or adverbs derived from these nouns and adjectives. *Example*: 'humility' | 14 (4%) | 2 (1%) |
| Other | This criterion serves as a residual category and contains all criteria which could not be assigned to any of the other evaluation criteria. *Example*: 'Letters of support are lacking.' | 7 (2%) | 7 (3%) |

*Note.* Percentages do not add up to 100% due to rounding.

The analysis of the overlap of criteria yielded a Jaccard Index of 0.39, which indicates weak overlap among the included studies when applying the benchmarks of Evans (1996) as suggested by Fried (2017). Originality was the only criterion which appeared in all studies. Other common criteria were clarity (10 studies), academic relevance (8 studies), extra-academic relevance (8 studies), and feasibility (8 studies). The least common criteria were motivation and traits, which appeared only in Lamont (2009) and van Arensbergen et al. (2014b). A complete breakdown of criteria by studies is provided in the Supplementary Materials (Part D).

*Association between criteria and entities.* The included studies almost never reported a criterion without a related entity (7 occurrences, 2% of the totally assigned entity codes; full counting, N = 373). In comparison, the studies reported entities without a related criterion often (48 occurrences, 13% of the totally assigned entity codes; full counting, N = 373). Among these entities without criteria, applicant entities were overrepresented (28 occurrences, 35% of the totally assigned applicant codes; full counting, n = 80) and social skills was the only entity for which no related criterion has been reported in any study. On average, the studies reported one criterion per entity (mode = 1, mean = 1.05; full counting, n = 360 after excluding 'no entity reported', n = 7, and 'other entity', n = 6).

The criteria were used in the following configuration to evaluate the main entities applicant, project, and environment. Originality, relevance (general, academic, non-academic), rigor, and coherence/justification were exclusively used to assess the project. Also exclusive were motivation and traits, which were only used to evaluate the applicant. In contrast, quality, clarity, and completeness were used with both project and applicant. Appropriateness, diversity, and feasibility were the only criteria which were used to assess all three main entities. Figure 2 provides a detailed overview of the association between criteria and sub-entities. It shows that, for example, the criteria appropriateness and completeness were used to evaluate 19 different sub-entities. In contrast, motivation and traits were used to assess just one sub-entity (applicant in general). On average, a criterion was used to evaluate 8.9 different sub-entities (median = 8, minimum = 1, maximum = 19).

A network analysis was conducted to examine the association between evaluation criteria and evaluated entities in terms of content. Using the the DIRTLPAwb+ algorithm, six communities of criteria and entities were detected (Figure 2). Since we could not interpret community 3 on its own in a meaningful way and since it is semantically related to community 4, we merged the two communities, which reduced the total number of communities to five. These five communities were included in the bipartite network (Figure 3) and are indicated by colors (e.g. red community). Accordingly, red corresponds to community 1 identified with the DIRTLPAwb+ algorithm, orange to community 2, green to the communities 3 and 4, blue to community 5, and purple to community 6.

**Figure 2**. Association of evaluation criteria and evaluated entities (conceptual counting). Association frequency is indicated in shades of blue. Communities detected with the DIRTLPAwb+ algorithm are marked red. Capital letters indicate to which main entity the sub-entities belong (A = applicant, P = project, E = environment).

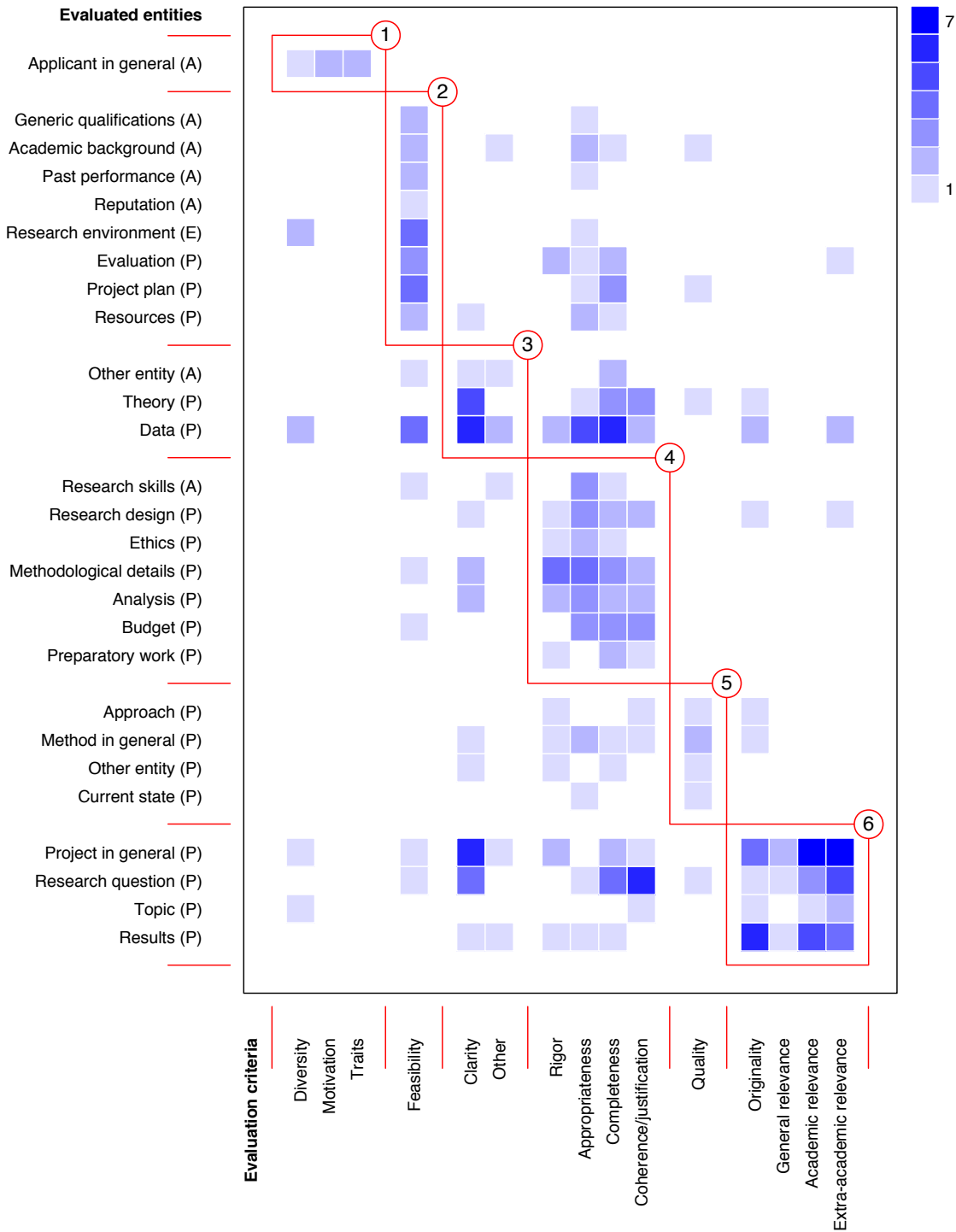

**Figure 3**. Bipartite network of evaluation criteria and evaluated entities (conceptual counting). Criteria are displayed in upper case (RIGOR) and entities in lower case (project plan). Node size indicates the frequency with which criteria and entities occurred. Edge size indicates association frequency. Communities of criteria and entities are indicated by colors.

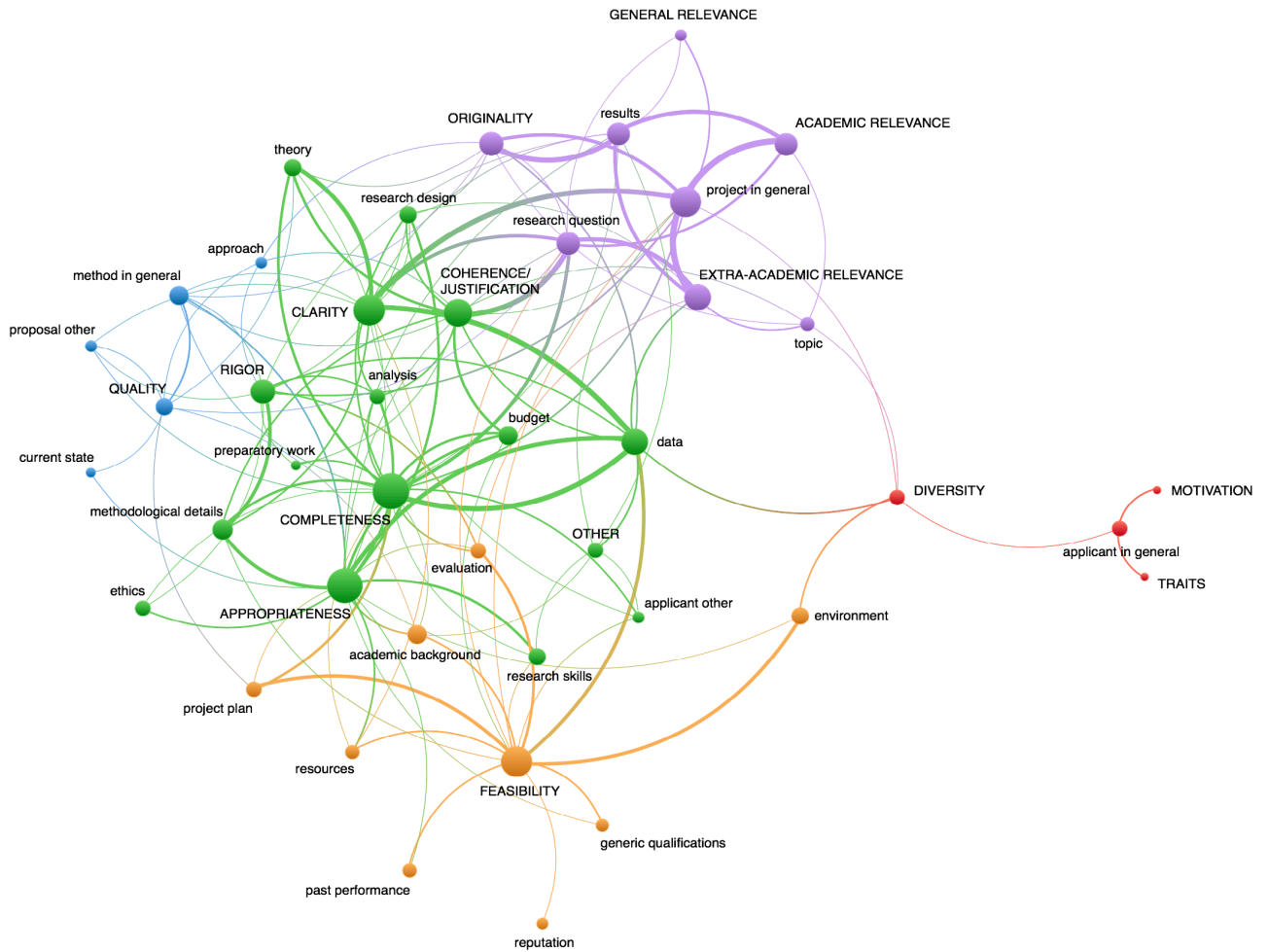

The communities portrayed in the bipartite network (Figure 3) can be described as follows. The *red community* is the smallest of all communities and comprises one general entity (applicant in general) and three criteria (motivation, traits, diversity). This community is only weakly connected to the other communities and focuses on assessing the person and personality of the applicant from a 'psychological' (motivation, traits) and a 'sociological' (diversity) perspective. More tangible aspects of the applicant, such as research skills and academic background are part of the orange and green communities. The *orange community* consists of one criterion (feasibility) and eight entities pertaining to the applicant, the project, and the environment. It is well connected to the green community, which is also reflected in an overlap of the two communities. In particular, research skills and budget are entities that may fit semantically better to entities in the orange than in the green community, but they are strongly tied to criteria in the green community (appropriateness, completeness, coherence/justification). Conversely, the entity evaluation may fit better to the entities in the green than in the orange community, but it is strongly tied to the criterion feasibility in the orange community. The orange community, including the overlap with the green community, suggests that the feasibility of a proposed project is assessed based on the qualifications (generic qualifications), achievements (past performance, reputation, academic background) and abilities (research skills) of the applicant, as well as on the available or requested resources (research environment, resources, budget), and the project plan. The *green community* is the largest community (six criteria, ten entities) and is closely connected to the orange, blue, and purple communities. Taking into account the overlap with the orange community, the focus of the green community is on assessing the proposed research process (preparatory work, theory, data, ethics, methodological details, research design, evaluation, analysis) on the content level in terms of rigor, appropriateness, and coherence/justification as well as on the descriptive level in terms of clarity and completeness. The *blue community* is closely connected to the green community and consists of one general evaluation criterion (quality) and four entities (method in general, approach, current state, proposal other). In contrast to the green community, here, the research process is assessed in more general terms. For example, while in the blue community the 'quality' of the 'method in general' is assessed, the green community evaluates the 'rigor' and 'appropriateness' of 'methodological details'. Lastly, the *purple community* is closely tied to the green community and includes four criteria (originality; general, academic, extra-academic relevance) and four entities (project in general, research question, topic, results). It focuses on assessing the originality and relevance of the starting point of the proposed project (research question, topic) as well as its endpoint (results).

*Conceptualization*. Based on the bipartite network and its communities, we derived an overall conceptualization of the evaluation criteria and evaluated entities involved in grant peer review. In this conceptualization, the criteria and entities are structured into *aims*, *means*, and *outcomes* (Figure 4). Thereby, the means describe how the aims are to be achieved in terms of the *research process* and the *project resources*. The aims and outcomes correspond to the purple community in the bipartite network (see Figure 3), the research process to the blue and green communities, and the project resources to the red and orange communities. Entities defined in general terms (e.g. project in general) and residual categories (e.g. other entity, other criteria) were not included in the conceptualization.

In our conceptualization, the *aims* comprise the research questions, hypotheses, goals, and the scope or topic of the proposed project. They are assessed in terms of originality, academic relevance, and extra-academic relevance. The *research process* includes those research steps and elements that are necessary to achieve the aims (e.g. preparatory work, data, theory, method, analysis). They are evaluated both on the content level (quality, appropriateness, rigor, coherence/justification) as well as on the level of description (clarity, completeness). The *project resources* include the resources needed to implement the research process, such as the requested budget or available equipment, facilities, and staff. They also comprise the project plan and timeline of the project. Moreover, in the included studies, the applicant is given an instrumental role in the implementation of the project and, therefore, she/he is represented as a project resource in Figure 4. The project resources, including the abilities and achievements of the applicant, are evaluated in terms of feasibility. The applicant's person and personality, however, are assessed in terms of motivation, traits, and diversity. The *outcomes* include the expected results of the proposed project as well as the anticipated benefits and consequences (e.g. 'outcome', 'impact', 'improvements') and are evaluated in terms of originality, academic relevance, and extra-academic relevance.

**Figure 4**. Conceptualization of evaluation criteria and evaluated entities used in grant peer review. Evaluated entities identified in the qualitative content analysis are displayed in boxes and regular type. Evaluation criteria are linked to evaluated entities by grey lines.

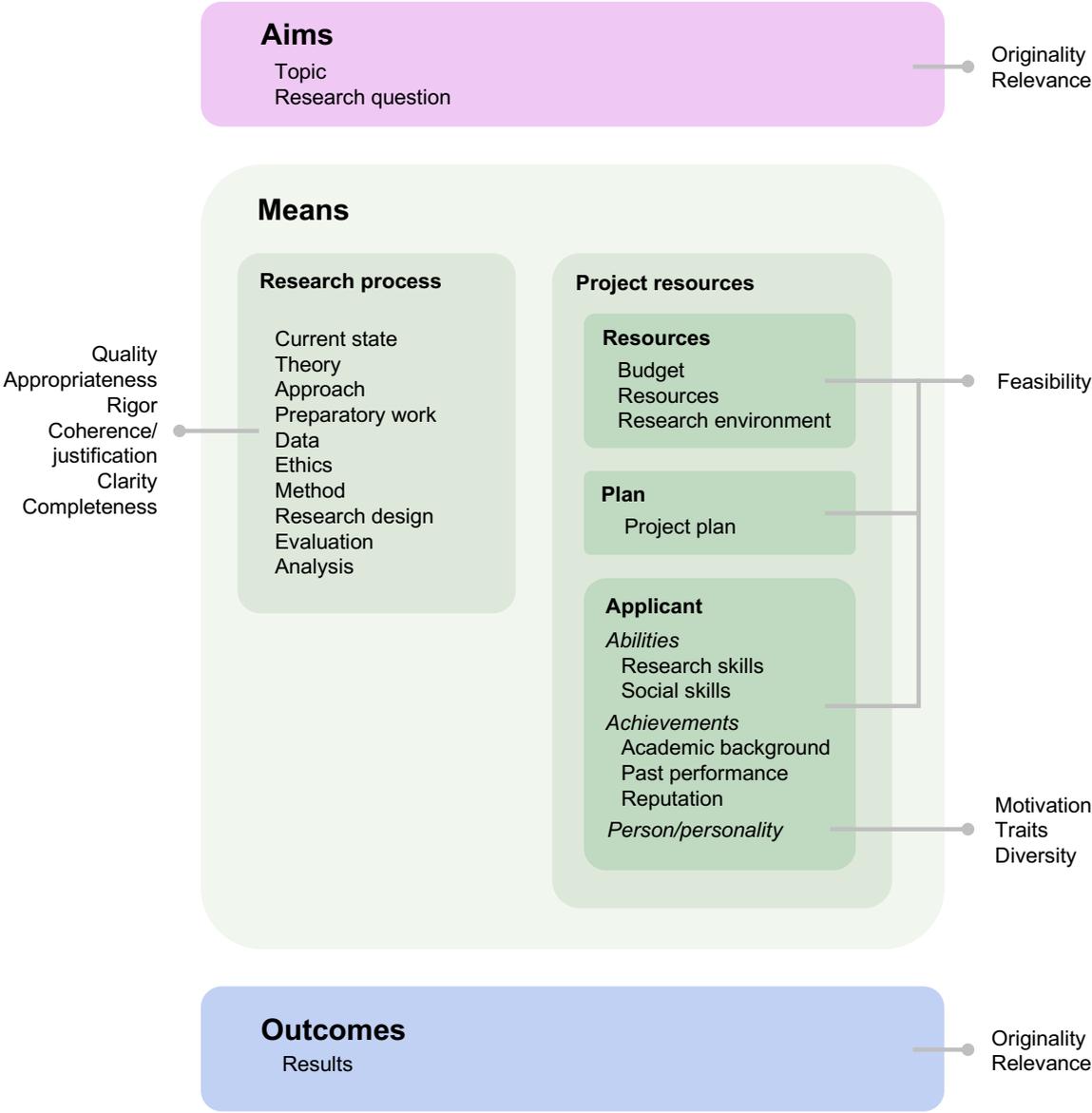

## Discussion and conclusion

In this article, we have synthesized 12 studies that examined grant peer review criteria in an empirical and inductive manner. To facilitate the synthesis, we introduced a framework that classifies what is generally referred to as 'criterion' into an *evaluated entity* (i.e. the object or target of the evaluation) and an *evaluation criterion* (i.e. the dimension along which an entity is evaluated). We conducted a qualitative content analysis of the 12 studies and thereby identified 15 evaluation criteria and 30 evaluated entities as well as the relations between them. Based on a network analysis, we proposed a conceptualization, which groups these evaluation criteria and evaluated entities into *aims*, *means*, and *outcomes*. In this last section, we compare our results to criteria found in studies on research quality and guidelines of funding agencies, discuss our results in relation to two normative positions, the fairness doctrine and the ideal of impartiality, and present limitations and avenues for future research.

Aksnes *et al.* (2019) argue from a context-independent perspective that originality, scientific value, societal value, and plausibility/soundness are the key dimensions of research quality and that each of these four dimensions includes a variety of aspects, which may be context-dependent. According to our analysis, these four dimensions are clearly present in grant peer review. While the first three dimensions match our evaluation criteria originality, academic relevance, and extra-academic relevance, the fourth dimension, plausibility/soundness, corresponds to a group of criteria, that is to appropriateness, rigor, coherence/justification, and quality. Since we have analyzed the association between evaluation criteria and evaluated entities, we can also indicate the specific entities mainly evaluated by these four dimensions of research quality. While originality, academic relevance, and extra-academic relevance were mostly used to assess the aims and the expected results of the proposed project, appropriateness, rigor, coherence/justification, and quality were mostly used to assess entities pertaining to the research process (e.g. data, theory, method, analysis). In addition to the four dimensions of Aksnes et al. (2019), our synthesis has shown that three other dimensions are important in assessing the merit of grant proposals. The first dimension, *quality of description*, assesses how the proposed project and information about the applicant are reported and presented in terms of the criteria clarity and completeness. The second dimension, *personal qualities*, assesses the person and personality of the applicant from a 'psychological' (motivation, traits) and a 'sociological' (diversity) perspective. Lastly, the resources needed to implement the project (e.g. project plan, budget, research environment, applicant's abilities) are evaluated in terms of the criterion *feasibility*. Based on these considerations, the criteria identified in this systematic review can be summarized as follows: evaluation criteria used by peers to assess grant applications = research quality (originality; academic and extra-academic relevance; quality, appropriateness, rigor, coherence/justification) + quality of description (clarity, completeness) + personal qualities (motivation, traits; diversity) + feasibility. This 'criteria formula' does not imply that peers use each evaluation criterion in every assessment. Rather, we understand it as a repertoire from which peers choose when assessing grant applications. In addition, we conceive the evaluation criteria (and evaluated entities) as situated concepts (Yeh and Barsalou 2006) that are (re)shaped by the actual assessment in which they are enacted (Kaltenbrunner and de Rijcke 2019).

Prescriptive criteria of funding agencies, as summarized by Abdoul et al. (2012), Berning et al. (2015), Falk-Krzesinski and Tobin (2015), and Langfeldt and Scordato (2016), generally overlap with the criteria of peers identified in this article in terms of research quality, quality of description, and feasibility. They differ, however, in two important respects. First, only the criteria of peers include an assessment of the applicant in terms of the personal qualities motivation and traits. Second, the criteria of peers do not include criteria emphasized by funding agencies, such as strategic importance (Berning et al. 2015), promotion of the public understanding of science (Abdoul et al. 2012), environmental sustainability (Langfeldt and Scordato 2016), or return on investment (Falk-Krzesinski and Tobin 2015). This supports Guston's (2000) view of funding agencies as 'boundary organizations' that stabilize the boundary between the research and the policy domain against external forces and, at the same time, continue to negotiate this boundary internally. The overlap of criteria of funding agencies with the criteria of peers outlined above is a sign of the stability of the boundary while the differences indicate that negotiations are continuing. Since our comparison of the criteria of peers and funding agencies is very general, future research should address this in more detail, for example, by using the framework of Langfeldt *et al.*' (2019), which is designed to study context-specific understandings of research quality.

Since peer review is often approached from a normative perspective, we discuss our findings in relation to two normative positions, the fairness doctrine and the ideal of impartiality. In their seminal work on peer review, Peters and Ceci (1982, p. 252) articulated the 'fairness doctrine' which holds that access to journal space and federal funds has to be 'judged on the merit of one's ideas, not on the basis of academic rank, sex, place of work, publication record, and so on'. The fairness doctrine resembles the ideal of impartiality, which implicitly underlies quantitative research on bias in peer review (Lee et al. 2013). The impartiality ideal requires that 'evaluative criteria have to do with the cognitive content of the submission' and reviewers have to 'interpret and apply evaluative criteria in the same way in the assessment of a submission' (Lee et al. 2013, pp. 3–4). In this way, evaluations are 'independent of the author's and reviewer's social identities and independent of the reviewer's theoretical biases and tolerance for risk' (Lee et al. 2013, p. 4). According to the fairness doctrine and the ideal of impartiality, grant peer review is unfair and biased because peers assess proposals, as our synthesis has shown, also in terms of non-epistemic criteria such as the applicant's reputation, past performance, academic background, skills, and personality. From the perspective of the fairness doctrine and the impartiality ideal, this biasedness implies that peer review should either be abolished or that non-epistemic components should be excluded from the assessment process. As peer review is regarded as indispensable in science and because epistemic and social dimensions are inseparable in peer review (Derrick 2018, Gläser and Laudel 2005, Hirschauer 2004, Lamont 2009, Lipworth *et al.* 2011, Reinhart 2012), we do not consider these options viable. Instead, we suggest following Lee et al. (2013), who proposed to develop alternative normative models, which acknowledge the sociality and partiality of peer review. We think that the philosophical debate on values in science (e.g. Douglas 2009, Elliott 2017) could prove to be particularly fruitful for this purpose as it started from the value-free ideal, which is similar to the fairness doctrine and impartiality ideal, and has advanced to acknowledging and including non-epistemic values. Drawing on Douglas (2016) and Elliott (2017), the following questions may guide the development of new normative models for peer review. What are the major ways in which values influence peer review? Which values are legitimate in peer review? When and how are they legitimate?[4]

Although this article cannot offer any recommendations for peer review practice from a normative perspective, it informs on which entities peers focus on when they assess grant proposals and along which criteria they assess these entities. This information may be useful for early career researchers who draft a grant proposal or learn to review, for the broader scientific community when discussing normative models for peer review, or for funding agencies that train reviewers. Moreover, this article provides a framework for analyzing assessment criteria, which may be useful for funders in setting up and revising their review criteria.

This article has the following main limitations. As there is no distinct discourse on the nature of (grant) peer review criteria in the literature, relevant studies were difficult to identify. In addition, our search strategy was English-language based. It is thus possible that not all pertinent studies are covered in this systematic review. Moreover, this systematic review does not adequately represent review practices in grant funding as the number of included studies is small and certain research fields (medical and health sciences), regions (USA, Europe), stages of the review process (individual review), and types of funding programs (project funding) are overrepresented. We therefore expect that future studies will discover additional evaluated entities and evaluation criteria. We also expect that future studies need to conceptualize the role of the applicant differently if they focus on criteria in scholarship and fellowship programs. According to our synthesis, which is mostly based on studies on project funding, peers evaluate the applicant in terms of a resource needed to implement the proposed project (instrumental role). In fellowship programs, however, peers evaluate the applicant also in order to decide whether she/he is suited for a further step in her/his academic career (Kaltenbrunner and de Rijcke 2019). Lastly, our findings depend on the choices made in the included studies, such as the number of reported criteria, the level of abstraction of the reported criteria, or the words used to describe the criteria. We assume that such choices could be the reason why we found a weak overlap of evaluation criteria among the included studies and why we found a high frequency of generally defined evaluated entities (e.g. 'project').

Future research on criteria in grant peer review should focus first and foremost on how exactly applicants are assessed as the studies included in this systematic review reported only few entities and criteria related to applicants. Second, because most of the included studies focused on individual reviews of written applications, other stages of the review process should be analyzed in future studies as well. In particular, we suggest examining the criteria employed in panel discussions and in interviews of applicants as these stages are central to grant peer review but rarely researched. Third, future studies should examine fields other than the medical and health sciences and include data from non-Western countries to test if criteria vary across disciplines and regions. Lastly, bias factors identified in quantitative studies on grant peer review should be summarized and added to the entities and criteria found in this systematic review to gain a comprehensive understanding of how peers consciously and unconsciously assess grant applications.

**Notes**

1. Generating evidence on the content validity of peer review can be a goal in itself, but content validation could also be a way to overcome the circularity inherent to the validation strategies discussed and employed in research on peer review and bibliometrics. For example, Moed (2005) und Harnad (2008, 2009) discuss the validation of bibliometric indicators by correlating them with peer review ratings (i.e. the criterion variable). Some studies also proceed conversely and seek to validate peer judgments with bibliometric indicators (i.e. the criterion variable in this case). These validation strategies require that the criterion variable has already been validated (AERA *et al.* 2014, Kane 2006) but neither peer review ratings nor bibliometric indicators fulfill this requirement (e.g. Cronin and Sugimoto 2015, Harnad 2008, Marsh *et al.* 2008). From our point of view, a possible solution could be to generate evidence on the content validity of peer review and thus validate peer ratings (for content validation, see Haynes *et al.* 1995).
2. The paradigm resembles what is described in the Standards for Educational Psychological Testing (AERA *et al.* 2014) and the Handbook of Test Development (Lane *et al.* 2016). It casts peer review as an instrument or test that has to be evaluated with respect to efficiency, reliability, fairness, and (predictive) validity.
3. Since this systematic review focuses on studies that have analyzed criteria applied by peers (i.e. descriptive-inductive studies), it is important to emphasize that there are three other types of studies on criteria (i.e. normative-theoretical, normative-prescriptive, and descriptive-deductive studies). These four types are described in the Supplementary Materials (Part A).
4. We understand a value as 'something that is desirable or worthy of pursuit' (Elliott 2017, p. 11). For example, the evaluation criterion 'originality' is an epistemic value while 'extra-academic relevance' is a social value.

# References

Abdoul H, Perrey C, Amiel P, Tubach F, Gottot S, Durand-Zaleski I and Alberti C (2012) Peer review of grant applications: Criteria used and qualitative study of reviewer practices. *PLOS ONE* 7(9): 1–15.

AERA, APA and NCME (2014) *Standards for educational and psychological testing.* American Educational Research Association: Washington.

Agarwal R, Chertow G M and Mehta R L (2006) Strategies for successful patient oriented research: Why did I (not) get funded? *Clinical Journal of the American Society of Nephrology* 1(2): 340–343.

Aksnes D W, Langfeldt L and Wouters P (2019) Citations, citation indicators, and research quality: An overview of basic concepts and theories. *SAGE Open* 9(1): 1–17.

Allen E M (1960) Why are research grant applications disapproved. *Science* 132(3439): 1532–1534.

Aagaard (2019). Quality criteria and concentration of research funding. Available at: https://www.r-quest.no/wp-content/uploads/2019/12/R-Quest_Policy_brief_4_2019.pdf [Accessed 19 Dec 2019].

Andersen J P (2013) *Conceptualising research quality in medicine for evaluative bibliometrics.* University of Copenhagen: Copenhagen.

Beckett S J (2016) Improved community detection in weighted bipartite networks. *Royal Society Open Science* 3(1): 1–18.

Belter C W (2016) Citation analysis as a literature search method for systematic reviews. *Journal of the Association for Information Science and Technology* 67(11): 2766–2777.

Berning N, Nünning A and Schwanecke C (2015) (Trans-)national criteria, norms and standards in literary studies: A comparative analysis of criteria-based ex ante evaluation forms of funding proposals in the humanities. *Germanisch-Romanische Monatsschrift* 65(1): 115–135.

Bootzin R R, Sechrest L, Scott A and Hannah M (1992) Common methodological problems in health services research proposals. *EGAD Quarterly* 1(3): 101–107.

Bornmann L and Daniel H D (2005) Selection of research fellowship recipients by committee peer review. Reliability, fairness and predictive Validity of board of trustees' decisions. *Scientometrics* 63(2): 297–320.

Bornmann L, Nast I and Daniel H D (2008) Do editors and referees look for signs of scientific misconduct when reviewing manuscripts? A quantitative content analysis of studies that examined review criteria and reasons for accepting and rejecting manuscripts for publication. *Scientometrics* 77(3): 415–432.

Boyack K W, Smith C and Klavans R (2018) Toward predicting research proposal success. *Scientometrics* 114(2): 449–461.

Chase J M (1970) Normative criteria for scientific publication. *The American Sociologist* 5(3): 262–265.

Chubin D E and Hackett E J (1990) *Peerless science. Peer review and U.S. science policy.* State University of New York Press: Albany.

Coveney J, Herbert D L, Hill K, Mow K E, Graves N and Barnett A (2017) Are you siding with a personality or the grant proposal? Observations on how peer review panels function. *Research Integrity and Peer Review* 2(19): 1–14.

Cronin B and Sugimoto C R (2015) *Scholarly metrics under the microscope. From citation analysis to academic auditing.* Information Today: New Jersey.

Cuca J M (1983) NIH grant applications for clinical research: Reasons for poor ratings or disapproval. *Clinical Research* 31(4): 453–463.

Davidson J (2005) Criteria. In: Mathison S (ed) *Encyclopedia of Evaluation*. SAGE: Thousand Oaks, pp 91–92.

Derrick G (2018) *The evaluators eye. Impact assessment and academic peer review.* Palgrave Macmillan: Cham.

Dormann C F, Gruber B and Fruend J (2008) Introducing the bipartite package: Analysing ecological networks. *R News* 8(2): 8–11.

Douglas H (2009) *Science, policy, and the value-free ideal.* University of Pittsburgh Press: Pittsburgh.

Douglas H (2016) Values in science. In: Humphreys P (ed) *The Oxford Handbook of Philosophy of Science*. Oxford University Press: New York, pp 609–632.

Elliott K C (2017) *A tapestry of values: An introduction to values in science.* Oxford University Press: Oxford.

European Science Foundation (2011) *European peer review guide. Integrating policies and practices into coherent procedures.* ESF: Strasbourg.

Evans J D (1996) *Straightforward statistics for the behavioral sciences.* Brooks/Cole Publishing: Pacific Grove, CA.

Falk-Krzesinski H J and Tobin S C (2015) How do I review thee? Let me count the ways: A comparison of research grant proposal review criteria across US federal funding agencies. *The Journal of Research Administration* 46(2): 79–94.

Fournier D M (1995) Establishing evaluative conclusions: A distinction between general and working logic. *New Directions for Evaluation* 1995(68): 15–32.

Fried E I (2017) The 52 symptoms of major depression: Lack of content overlap among seven common depression scales. *Journal of Affective Disorders* 208: 191–197.

Fuller E O, Hasselmeyer E G, Hunter J C, Abdellah F G and Hinshaw A S (1991) Summary statements of the NIH nursing grant applications. *Nursing Research* 40(6): 346–351.

Gesuato S (2009) Evaluation guidelines: A regulatory genre informing reviewing practices. In: Gotti M (ed) *Commonality and Individuality in Academic Discourse*. Peter Lang: Bern, pp 325–348.

Gläser J and Laudel G (2005) Advantages and dangers of 'remote' peer evaluation. *Research Evaluation* 14(3): 186–198.

Goertz G (2006) *Social science concepts. A user's guide.* Princeton University Press: Princeton.

Gough D (2007) Weight of evidence: A framework for the appraisal of the quality and relevance of evidence. *Research Papers in Education* 22(2): 213–228.

Gough D (2015) Qualitative and mixed methods in systematic reviews. *Systematic Reviews* 4(181): 1–3.

Gregorius S, Dean L, Cole D C and Bates I (2018) The peer review process for awarding funds to international science research consortia: A qualitative developmental evaluation. *F1000Research* 6(1808): 1–22.

Guetzkow J, Lamont M and Mallard G (2004) What is originality in the humanities and the social sciences? *American Sociological Review* 69(2): 190–212.

Gulbrandsen M J (2000) *Research quality and organisational factors: An investigation of the relationship.* Norwegian University of Science and Technology: Trondheim.

Guston D H (2000) *Between politics and science: Assuring the integrity and productivity of research*. Cambridge University Press: New York.

Guthrie S, Ghiga I and Wooding S (2018) What do we know about grant peer review in the health sciences? [version 2; referees: 2 approved]. *F1000Research* 6(1335): 1–23.

Guthrie S, Rodriguez Rincon D, MacInroy G, Ioppolo B and Gunashekar S (2019) Measuring bias, burden and conservatism in research funding processes [version 1; peer review: 1 approved, 1 approved with reservations]. *F1000Research* 8(851): 1–27.

Harnad S (2008) Validating research performance metrics against peer rankings. *Ethics in Science and Environmental Politics* 8:103–107.

Harnad S (2009) Open access scientometrics and the UK Research Assessment Exercise. *Scientometrics* 79(1): 147–156.

Hartmann I (1990) *Begutachtung in der Forschungsförderung: Die Argumente der Gutachter in der Deutschen Forschungsgemeinschaft.* RG Fischer: Frankfurt.

Hartmann I and Neidhardt F (1990) Peer review at the Deutsche Forschungsgemeinschaft. *Scientometrics* 19(5-6): 419–425.

Haynes S N, Richard D and Kubany E S (1995) Content validity in psychological assessment: A functional approach to concepts and methods. *Psychological Assessment* 7(3): 238–245.

Hemlin S (1993) Scientific quality in the eyes of the scientist: A questionnaire study. *Scientometrics* 27(1): 3–18.

Hemlin S and Montgomery H (1990) Scientists' conceptions of scientific quality: An interview study. *Science Studies* 3(1): 73–81.

Hemlin S and Montgomery H (1993) Peer judgments of scientific quality: A cross-disciplinary document analysis of professorship candidates. *Science Studies* 6(1): 19–27.

Hemlin S, Niemenmaa P and Montgomery H (1995) Quality criteria in evaluations: Peer reviews of grant applications in psychology. *Science Studies* 8(1): 44–52.

Hewings M (2004) An important contribution or tiresome reading? A study of evaluation in peer reviews of journal article submissions. *Journal of Applied Linguistics* 1(3): 247–274.

Hirschauer S (2004) Peer Review Verfahren auf dem Prüfstand. *Zeitschrift für Soziologie* 33(1): 62–83.

Hug S E, Ochsner M and Daniel H D (2013) Criteria for assessing research quality in the humanities: A Delphi study among scholars of English literature, German literature and art history. *Research Evaluation* 22(5): 369–383.

Hume K M, Giladi A M and Chung K C (2015) Factors impacting successfully competing for research funding: An analysis of applications submitted to the plastic surgery foundation. *Plastic and Reconstructive Surgery* 135(2): 429–435.

International Cake Exploration Societé (2007). Guidelines for cake show judging. Available at: https://ucanr.edu/sites/4-H_Tulare/files/161715.pdf [Accessed 6 Dec 2018].

Johnson D R and Hermanowicz J C (2017) Peer Review: From 'sacred ideals' to 'profane realities' In: Paulsen M B (ed) *Higher Education: Handbook of Theory and Research.* Springer: Cham, Switzerland, pp 485–527.

Kaatz A, Magua W, Zimmerman D R and Carnes M (2015) A quantitative linguistic analysis of National Institutes of Health R01 application critiques from investigators at one institution. *Academic Medicine* 90(1): 69–75.

Kaltenbrunner W and de Rijcke S (2019) Filling in the gaps: The interpretation of curricula vitae in peer review. *Social Studies of Science*.

Kane M (2006) Content-related validity evidence in test development. In: Downing S M and Haladyna T M (eds) *Handbook of test development.* Lawrence Erlbaum: Mahwaw, pp 131-153.

Krippendorff K (2004) *Content analysis: An introduction to Its methodology.* 2nd edn. Sage: Thousand Oaks, CA.

Krippendorff K (2011). Computing Krippendorff's alpha-reliability. Available at: https://repository.upenn.edu/asc_paper/43/ [Accessed 6 Dec 2018].

Kuhn T S (1977). *The essential tension. Selected studies in scientific tradition and change.* University of Chicago Press: Chicago.

Lahtinen E, Koskinen-Ollonqvist P, Rouvinen-Wilenius P, Tuominen P and Mittelmark M B (2005) The development of quality criteria for research: A Finnish approach. *Health Promotion International* 20(3): 306–315.

Lane S, Raymond M R and Haladyna T M (2016) *Handbook of test development*. Routledge: New York.

Lamont M (2009) *How professors think: Inside the curious world of academic judgment.* Harvard University Press: Cambridge.

Lamont M and Guetzkow J (2016) How quality is recognized by peer review panels: The case of the humanities. In: Ochsner M, Hug S E and Daniel H D (eds) *Research Assessment in the Humanities. Towards Criteria and Procedures*. SpringerOpen: Cham, pp 31–41.

Landis J R and Koch G G (1977) The measurement of observer agreement for categorical data. *Biometrics* 33(1): 159–174.

Langfeldt L (2001) The decision-making constraints and processes of grant peer review, and their effects on the review outcome. *Social Studies of Science* 31(6): 820–841.

Langfeldt L, Nedeva M, Sörlin S and Thomas D A (2019) Co-existing notions of research Quality: A framework to study context-specific understandings of good research. *Minerva* advance online publication 26 August, doi: 10.1007/s11024-019-09385-2.

Langfeldt L and Scordato L (2016) *Efficiency and flexibility in research funding. A comparative study of funding instruments and review criteria.* Nordic Institute for Studies in Innovation, Research and Education: Oslo.

Lee C J, Sugimoto C R, Zhang G and Cronin B (2013) Bias in peer review. *Journal of the American Society for Information Science and Technology* 64(1): 2–17.

Lipworth W L, Kerridge I H, Carter S M and Little M (2011) Journal peer review in context: A qualitative study of the social and subjective dimensions of manuscript review in biomedical publishing. *Social Science & Medicine* 72(7): 1056–1063.

Marsh H W, Jayasinghe U W and Bond N W (2008) Improving the peer-review process for grant applications. *American Psychologist* 63(3): 160–168.

Mårtensson P, Fors U, Wallin S B, Zander U and Nilsson G H (2016) Evaluating research: A multidisciplinary approach to assessing research practice and quality. *Research Policy* 45(3): 593–603.

Mayring P (2014) *Qualitative content analysis: Theoretical foundation, basic procedures and software Solution.* Klagenfurt.

Mayring P (2015) *Qualitative Inhaltsanalyse. Grundlagen und Techniken.* 12th edn. Beltz: Weinheim/Basel.

Meierhofer L (1983) *Projektselektion in der Forschungsförderung.* Verlag Paul Haupt: Bern.

Merton R K (1973) *The Sociology of Science. Theoretical and empirical investigations*. University of Chicago Press: Chicago.

Moed H F (2005) *Citation analysis in research evaluation*. Springer: Dordrecht.

Moghissi A A, Love B R and Straja S R (2013) *Peer review and scientific Assessment. A handbook for funding organizations, regulatory agencies, and editors.* Institute for Regulatory Science: Alexandria.

Montgomery H and Hemlin S (1991) Judging scientific quality. A cross-disciplinary investigation of professorial evaluation documents. *Göteborg Psychological Reports* 21(4): 1–36.

Moore F D (1961) The surgery study section of the National Institutes of Health. *Annals of Surgery* 153(1): 1–12.

Mow K E (2011) Peers inside the black box: Deciding excellence. *International Journal of Interdisciplinary Social Sciences* 5(10): 175–184.

Neidhardt F (1988) *Selbststeuerung in der Forschungsförderung. Das Gutachterwesen der DFG.* Westdeutscher Verlag: Opladen.

NWO (2017a). International peer-review conference. Main outcomes. Available at: https://www.nwo.nl/binaries/content/documents/nwo-en/common/documentation/application/nwo/policy/main-outcomes-nwo-international-peer-review-conference-2017/Main-outcomes-NWO-international-peer-review-conference-oct-2017-pdf.pdf [Accessed 17 Sept 2019].

NWO (2017b). NWO measures to reduce application pressure. Available at: https://www.nwo.nl/binaries/content/documents/nwo-en/common/documentation/application/nwo/strategy/measures-to-reduce-application-pressure/NWO-measures-to-reduce-application-pressure_2017-pdf.pdf [Accessed 17 Sept 2019].

Ochsner M, Hug S E and Daniel H D (2013) Four types of research in the humanities: Setting the stage for research quality criteria in the humanities. *Research Evaluation* 22(2): 79–92.

OECD (2015) *Frascati Manual 2015: Guidelines for collecting and reporting data on research and experimental development. The Measurement of Scientific, Technological and Innovation Activities* OECD Publishing: Paris.

OECD (2018) *Effective operation of competitive research funding systems*. OECD Publishing: Paris.

Oortwijn W J, Vondeling H, van Barneveld T, van Vugt C and Bouter L M (2002) Priority setting for health technology assessment in the Netherlands: Principles and practice. *Health Policy* 62(3): 227–242.

Peters D P and Ceci S J (1982) Peer review practices of psychological journals: The fate of accepted, published articles, submitted again. *Behavioral and Brain Sciences* 5(2): 187–255.

Pier E L, Brauer M, Filut A, Kaatz A, Raclaw J, Nathan M J, Ford C E and Carnes M (2018) Low agreement among reviewers evaluating the same NIH grant applications. *Proceedings of the National Academy of Sciences of the United States of America* 115(12): 2952–2957.

Polanyi M (1962) The republic of science: Its political and economic theory. *Minerva* 1(1): 54–73.

Pollitt F A, Notgrass C M and Windle C (1996) Peer review of rural research grant applications. *Administration and Policy in Mental Health* 24(2): 173–180.

Porter A L and Rossini F A (1985) Peer review of interdisciplinary research proposals. *Science Technology & Human Values* 10(3): 33–38.

Prpić K and Šuljok A (2009) How do scientists perceive scientific quality. In: Prpić K (ed) *Beyond the Myths about the Natural and Social Sciences: A Sociological View*. Institute for Social Research: Zagreb, pp 205–245.

Publons (2019). Grant review in focus. Available at: https://publons.com/community/gspr/grant-review [Accessed 25 Oct 2019].

R Core Team (2018). R: A language and environment for statistical computing. R Foundation for Statistical Computing, Vienna. Available at: https://www.R-project.org.

Reinhart M (2010) Peer review practices: A content analysis of external reviews in science funding. *Research Evaluation* 19(5): 317–431.

Reinhart M (2012) *Soziologie und Epistemologie des Peer Review.* Nomos: Baden-Baden.

Research England, SFC, HEFCW and DfE (2018). REF 2021: Draft guidance on submissions. Available at: https://www.ref.ac.uk/media/1016/draft-guidance-on-submissions-ref-2018_1.pdf [Accessed 6 Dec 2018].

Sabaj Meruane O, González Vergara C and Pina-Stranger A (2016) What we still don't know about peer review. *Journal of Scholarly Publishing* 47(2): 180–212.

Schmitt J, Petzold T, Nellessen-Martens G and Pfaff H (2015) Prioritization and consentation of criteria for the appraisal, funding and evaluation of projects from the German

Innovation Fund: A multi-perspective Delphi study. *Gesundheitswesen* 77(8-9): 570–579.
Scriven M (1980) *The logic of evaluation.* Edgepress: Inverness.
Shadish W R (1989) The perception and evaluation of quality in science. In: Gholson B, Shadish W R, Neimeyer R A and Houts A C (eds) *Psychology in Science. Contributions to Metascience.* Cambridge University Press: New York, pp 382–426.
Shepherd J, Frampton G K, Pickett K and Wyatt J C (2018) Peer review of health research funding proposals: A systematic map and systematic review of innovations for effectiveness and efficiency. *PLOS ONE* 13(5): 1–26.
Staudt A (2019). icr: Compute Krippendorff's alpha. R package version 0.5.9. Available at: https://cran.r-project.org/web/packages/icr/ [Accessed 15 Feb 2019].
Tissot F, Hering S and Kleinberger U (2015) On-site visit interviews in external quality assurance procedures: A linguistic, empirical approach. In: Quality Assurance Agency and UCL Institute of Education, *10th European Quality Assurance Forum*. London, 19–21 November 2015.
Thomas J P and Lawrence T S (1991) Common deficiencies of NIDRR research applications. *American Journal of Physical Medicine & Rehabilitation* 70(1): 161–164.
Thorngate W, Dawes R M and Foddy M (2009) *Judging merit.* Taylor & Francis: New York.
Tong A, Flemming K, McInnes E, Oliver S and Craig J (2012) Enhancing transparency in reporting the synthesis of qualitative research: ENTREQ. *BMC Medical Research Methodology* 12(181): 1–8.
van Arensbergen P and van den Besselaar P (2012) The selection of scientific talent in the allocation of research grants. *Higher Education Policy* 25(3): 381–405.
van Arensbergen P, van der Weijden I and van den Besselaar P (2014a) The selection of talent as a group process. A literature review on the social synamics of decision making in grant panels. *Research Evaluation* 23(4): 298–311.
van Arensbergen P, van der Weijden I and van den Besselaar P (2014b) Different views on scholarly talent: What are the talents we are looking for in science? *Research Evaluation* 23(4): 273–284.
van den Besselaar P and Sandström U (2017) Influence of cognitive distance on grant decisions. *Science, Technology and Innovation Indicators 2017* Paris.
van den Besselaar P, Sandstrom U and Schiffbaenker H (2018) Studying grant decision-making: A linguistic analysis of review reports. *Scientometrics* 117(1): 313–329.
van den Broucke S, Dargent G and Pletschette M (2012) Development and assessment of criteria to select projects for funding in the EU health programme. *European Journal of Public Health* 22(4): 598–601.
Van Eck N J and Waltman L (2010) Software survey: VOSviewer, a computer program for bibliometric mapping. *Scientometrics* 84(2): 523–538.
Wenneras C and Wold A (1997) Nepotism and sexism in peer review. *Nature* 387(6631): 341–343.
Whaley A L, Rodriguez R and Alexander L A (2006) Development of a rating form to evaluate grant applications to the Hogg Foundation for Mental Health. *Evaluation Review* 30(1): 3–26.
White H D (2005) On extending informetrics: An opinion paper. In: Ingwersen P and Larsen B (eds) *Proceedings of the 10th International Conference of the International Society for Scientometrics and Informetrics*. Karolinska University Press: Stockholm, pp 442–449.
White H D (2016) Bibliometrics, librarians, and bibliograms. *Education for Information* 32(2): 125–148.
Yeh W C and Barsalou L W (2006) The situated nature of concepts. *American Journal of Psychology* 119(3): 349-384.

# Supplementary Materials

# Part A: Types of studies on criteria

In our view, studies on criteria relevant for grant peer review can be divided into the following four types. (1) Studies with a normative-theoretical focus discuss criteria and their normativity from a theoretical perspective. Examples include Polanyi's (1962) *Republic of Science*, Merton's (1973) *Ethos of Science*, and Kuhn's (1977) *Objectivity, Value Judgment, and Theory Choice*. (2) Studies with a normative-prescriptive focus categorize criteria specified by funding agencies (e.g. in guidelines for applicants and reviewers or in review forms). To our knowledge, there are four such studies (i.e. Abdoul et al. 2012, Berning et al. 2015, Falk-Krzesinski and Tobin 2015, Langfeldt and Scordato 2016). (3) Descriptive-deductive studies use theoretically derived factors to analyze actual assessments of and decisions on grant proposals with quantitative methods. According to Boyack et al. (2018), such studies focus on bias and fairness factors, such as gender (Wenneras and Wold 1997), and rarely address other factors, such as the scientific performance of applicants (Bornmann and Daniel 2005) or the cognitive distance between reviewers and applicants (van den Besselaar and Sandström 2017). Bias and fairness factors are often discussed in literature reviews (e.g. Guthrie et al. 2018, Lee et al. 2013), however, they are not referred to as criteria or contextualized as such. (4) Lastly, descriptive-inductive studies investigate criteria peers apply or deem appropriate. Such studies proceed inductively and frequently employ qualitative data analysis. For example, Lamont (2009) interviewed panelists on the criteria they have applied and Reinhart (2010) extracted criteria evident in written grant reviews.

# Part B: Analytical framework

**Example**

Based on the four components of the analytical framework, an evaluative act can be characterized as assigning a value (component d) to an evaluated entity (component a) by comparing the entity to the frame of reference (component c) of an evaluation criterion (component b). A result of an evaluative act may read like this: 'The buttercream icing looks flawless.' Based on the analytical framework and the Guidelines for Cake Show Judging (International Cake Exploration Societé 2007), this evaluative statement can be described as follows. Evaluated entity: cake covering (here: buttercream icing); evaluation criterion: appearance; assigned value: flawless; frame of reference:

judge's standard, probably ranging from 'needs improvement' (cake showing through, crumbs visible) to 'flawless' (smooth texture, no air bubbles, no streaks).

The example above comprises just one entity. However, according to the guidelines of the International Cake Exploration Societé (2007), there are four main entities that a cake show judge has to assess: the covering, the icing flowers, the technique, and the painting. Accordingly, the entity to be evaluated in a cake show could be conceived as a two-level concept with 'cake' constituting the basic level and the second level constituted by the four dimensions 'covering', 'icing flowers', 'technique', and 'painting'. Naturally, the substantive content of 'cake' could be extended and specified by adding further levels and dimensions.

**Unifying properties of the framework**

Components similar to those of our analytical framework can be found in studies on peer review criteria. For example, Hewings (2004), Gesuato (2009), and Langfeldt and Scordato (2016) use entities and criteria for structuring purposes and Hartmann and Neidhardt (1990) utilize criteria and values, but the terminology in these studies is, of course, different from our framework. Bornmann *et al.* (2008) also use criteria and values and additionally employ levels and dimensions to structure their findings. Hence, with our analytical framework it is possible to unify different structuring principles and terminologies that studies on peer review criteria have created ad hoc. The framework can also subsume Hemlin's conceptual system, which was developed from studies on peer review criteria (Hemlin and Montgomery 1990), extended to analyze reviews of applications for faculty positions (Montgomery and Hemlin 1991), and applied in several studies (Hemlin 1993; Hemlin and Montgomery 1993; Hemlin *et al.* 1995; Prpić and Šuljok 2009). Hemlin's system consists of four categories: the object of the judgement (e.g. journal article or grant application), the aspects of the research judged (e.g. method, theory, results), the attribute associated with the aspects (e.g. novelty, stringency), and the value of the attribute (e.g. positive, negative, neutral). Hemlin's attribute and value correspond to evaluation criterion and assigned value in our framework. Furthermore, Hemlin's object and aspects can be represented as a two-level entity in the framework, with the object as the basic level and aspects as the second level.

# Part C: Search strings

**Web of Science**

TS=((assess* OR evaluat* OR review* OR criteri* OR reject* OR *approve) AND ("research proposal*" OR "grant proposal*" OR "grant applica*" OR "grant allocat*" OR "grant panel*" OR "grant peer review*" OR "funding proposal*" OR "funding decision*" OR "research funding*" OR ("project funding*" NEAR research*) OR ("project funding*" NEAR scien*) OR ("project funding*" NEAR academ*) OR ("project selection*" NEAR research*) OR ("project selection*" NEAR scien*) OR ("project selection*" NEAR academ*) OR ("fellowship" NEAR research*) OR ("fellowship" NEAR scien*) OR ("fellowship" NEAR academ*) OR "scien*funding" OR "research*funding" OR "academ*funding"))

**Scopus**

TITLE-ABS-KEY((assess* OR evaluat* OR review* OR criteri* OR reject* OR *approve) AND ("research proposal*" OR "grant proposal*" OR "grant applica*" OR "grant allocat*" OR "grant panel*" OR "grant peer review*" OR "funding proposal*" OR "funding decision*" OR "research funding*" OR ("project funding*" W/15 research*) OR ("project funding*" W/15 scien*) OR ("project funding*" W/15 academ*) OR ("project selection*" W/15 research*) OR ("project selection*" W/15 scien*) OR ("project selection*" W/15 academ*) OR ("fellowship" W/15 research*) OR ("fellowship" W/15 scien*) OR ( "fellowship" W/15 academ*) OR "scien*funding" OR "research*funding" OR "academ*funding"))

## Part D: Overlap of evaluation criteria

Supplementary Figure S1. Evaluation criteria identified in the qualitative content analysis per included study.

| Study \ Criterion | Originality | Clarity | Acad. relevance | Extra-acad. relevance | Feasibility | Rigor | Appropriateness | Completeness | Coherence/justification | Quality | General relevance | Other | Diversity | Motivation | Traits | Total |
|---|---|---|---|---|---|---|---|---|---|---|---|---|---|---|---|---|
| Guetzkow et al. (2004), Lamont (2009) | ■ | ■ | ■ | ■ | ■ | ■ |  | ■ | ■ | ■ |  |  | ■ | ■ | ■ | 12 |
| Pollitt et al. (1996) | ■ | ■ | ■ | ■ | ■ | ■ | ■ | ■ | ■ |  |  | ■ | ■ |  |  | 11 |
| Schmitt et al. (2015) | ■ | ■ | ■ | ■ | ■ | ■ | ■ | ■ | ■ |  |  |  | ■ |  |  | 10 |
| Lahtinen et al. (2005) | ■ | ■ | ■ | ■ | ■ | ■ | ■ | ■ | ■ |  |  |  |  |  |  | 9 |
| Whaley et al. (2006) | ■ | ■ | ■ | ■ | ■ | ■ | ■ | ■ |  |  |  |  | ■ |  |  | 9 |
| Hartmann & Neidhardt (1990) | ■ | ■ | ■ | ■ | ■ |  | ■ |  | ■ | ■ |  |  |  |  |  | 8 |
| van Arensbergen et al. (2014b) | ■ | ■ |  |  | ■ |  |  |  | ■ | ■ | ■ |  |  | ■ | ■ | 8 |
| Abdoul et al. (2012) | ■ | ■ | ■ | ■ | ■ |  | ■ |  |  |  | ■ |  |  |  |  | 7 |
| Reinhart (2010) | ■ | ■ | ■ | ■ | ■ |  |  |  |  |  | ■ | ■ |  |  |  | 7 |
| Pier et al. (2018) | ■ | ■ |  |  |  |  | ■ |  |  |  | ■ |  |  |  |  | 4 |
| Thomas et al. (1991) | ■ |  |  |  |  | ■ |  |  |  | ■ |  | ■ |  |  |  | 4 |
| Total | 11 | 10 | 8 | 8 | 8 | 7 | 7 | 6 | 5 | 4 | 4 | 4 | 3 | 2 | 2 | 89 |

Using all 15 evaluation criteria to determine the overlap of the included studies yields a Jaccard Index of 0.39, which indicates weak overlap according to the benchmarks of Evans (1996). When merging general relevance, academic relevance, and extra-academic relevance and excluding the criterion 'other', the Jaccard Index slightly increases to 0.46, which indicates moderate overlap according to Evans' benchmarks.